\documentclass[aps,prd,preprint,tightenlines,superscriptaddress]{revtex4}
\usepackage{epsfig}
\usepackage{graphicx}
\usepackage{psfrag}
\usepackage{amsmath,amssymb}
\usepackage{colordvi}
\usepackage{color}
\usepackage{amsfonts}
\usepackage{enumerate}
\usepackage{slashed,bm}

\def \non {\nonumber}
\def \beq  {\begin{equation}}
\def \eeq  {\end{equation}}

\begin{document}

\title{One-Loop Matching for Generalized Parton Distributions}
\author{Xiangdong Ji}
\affiliation{INPAC, Department of Physics and Astronomy, Shanghai Jiao Tong University, Shanghai, 200240, People's Republic of China}
\affiliation{Center for High-Energy Physics, Peking University, Beijing, 100080, People's Republic of China}
\affiliation{Maryland Center for Fundamental Physics, Department of Physics,  University of Maryland, College Park, Maryland 20742, USA}
\author{Andreas Sch\"afer}
\affiliation{Institut f\"ur Theoretische Physik, Universit\"at Regensburg, \\
D-93040 Regensburg, Germany}
\author{Xiaonu Xiong}
\affiliation{Istituto Nazionale di Fisica Nucleare, Sezione di Pavia, Pavia 27100, Italy}
\affiliation{Center for High-Energy Physics, Peking University, Beijing 100080, People's Republic of China}
\author{Jian-Hui Zhang}
\affiliation{INPAC, Department of Physics and Astronomy, Shanghai Jiao Tong University, Shanghai, 200240, People's Republic of China}
\affiliation{Institut f\"ur Theoretische Physik, Universit\"at Regensburg, \\
D-93040 Regensburg, Germany}
\vspace{0.5in}
\begin{abstract}

We present the one-loop matching condition for the unpolarized and polarized generalized quark distributions in the nonsinglet case. The matching condition links the quasi distributions defined in terms of spacelike correlators at finite nucleon momentum to the light cone distributions, and it is useful for extracting the latter from the former in a lattice QCD calculation. Our results show that at one-loop and leading power accuracy the matching for the light cone generalized quark distribution $H$ ($\tilde H$) is nontrivial, whereas no matching is required for $E$ ($\tilde E$). Therefore, $E$ ($\tilde E$) can be smoothly approached by its quasi counterpart in the large momentum limit. We also present the matching for the distribution amplitude of the pion.

\end{abstract}

\maketitle

\section{introduction}
One of the important goals of quantum chromodynamics (QCD) is to understand the internal structure of nucleons in terms of the fundamental degrees of freedom of QCD -quarks and gluons. The parton distribution functions (PDFs) play a crucial role in characterizing the nucleon structure. They are defined as the forward hadronic matrix elements of light cone correlations, and they describe the momentum distributions of quarks and gluons inside the nucleon. In recent years, their generalization to nonforward kinematics, known as generalized parton distributions (GPDs)~\cite{Ji:1996ek, Radyushkin:1996nd, Ji:1996nm, Radyushkin:1996ru}, also received considerable attention (for recent reviews on GPDs see e.g.~\cite{Belitsky:2005qn,Ji:2004gf, Diehl:2003ny}). In contrast to parton distributions, the GPDs encode more information about the internal structure of nucleons, and can be viewed as a hybrid of parton distributions, form factors and distribution amplitudes. They played an important role in providing a three-dimensional spatial picture of the nucleon~\cite{Burkardt:2000za} and in revealing the spin structure of the nucleon~\cite{Ji:1996ek}. Experimentally, GPDs can be accessed in exclusive processes such as deeply virtual Compton scattering or meson production. However, defined as nonlocal light cone correlations, they are rather difficult to access by lattice QCD simulations.

Recently, a direct approach to accessing parton distributions and related quantities has been proposed~\cite{Ji:2013fga, Ji:2013dva, Hatta:2013gta, Xiong:2013bka, Lin:2014zya, Ma:2014jla, Ji:2014gla, Ji:2014hxa, Ji:2014lra, Ji:2015jwa,Alexandrou:2015rja}. According to this approach, the light cone parton distribution can be studied by investigating the large momentum limit of a quasi parton distribution, which is a time-independent spacelike correlation and thus can be simulated on a Euclidean lattice. The light cone distribution is then recovered from the quasi one by a factorization formula or matching condition. This procedure in principle applies not only to parton distributions, but also to other quantities defined on the light cone. In Ref.~\cite{Xiong:2013bka}, we presented a factorization formula connecting the light cone and quasi parton distributions and proved its validity up to one-loop order, where we showed that the quasi and light cone parton distributions have the same collinear singularities, and the matching factor connecting them is sensitive to UV physics only.

In this paper, we consider the one-loop matching for GPDs. In particular, we focus on the unpolarized GPDs $H(x,\xi,t)$ and $E(x,\xi,t)$, and the polarized ones $\tilde H(x, \xi, t)$ and $\tilde E(x, \xi, t)$, which are defined in terms of the following matrix elements
\begin{align}
F_q(x,\xi,t)&=\int\frac{dz^-}{4\pi}e^{ixp^+ z^-}\langle p''|\bar\psi(-\frac{z}{2})\gamma^+ L(-\frac{z}{2},\frac{z}{2})\psi(\frac{z}{2})|p'\rangle_{z^+=0,\vec z_\perp=0}\non\\
&=\frac{1}{2p^+}\big[H(x,\xi,t)\bar u(p'')\gamma^+ u(p')+E(x,\xi,t)\bar u(p'')\frac{i\sigma^{+\nu}\Delta_\nu}{2m}u(p')\big], \non\\
\tilde F_q(x,\xi,t)&=\int\frac{dz^-}{4\pi}e^{ixp^+ z^-}\langle p''|\bar\psi(-\frac{z}{2})\gamma^+ \gamma^5 L(-\frac{z}{2},\frac{z}{2})\psi(\frac{z}{2})|p'\rangle_{z^+=0,\vec z_\perp=0}\non\\
&=\frac{1}{2p^+}\big[\tilde H(x,\xi,t)\bar u(p'')\gamma^+ \gamma^5 u(p')+\tilde E(x,\xi,t)\bar u(p'')\frac{\gamma^5\Delta^+}{2m}u(p')\big],
\end{align}
where $L(-\frac{z}{2},\frac{z}{2})$ is the gauge link along the light cone and
\beq
p^\mu=\frac{p''^\mu+p'^\mu}{2},\hspace{2em} \Delta^\mu=p''^\mu-p'^\mu, \hspace{2em} t=\Delta^2, \hspace{2em} \xi=\frac{p''^+-p'^+}{p''^+ +p'^+}.
\eeq
In the limit $\xi, t\to 0$, $H$ and $\tilde H$ reduce to the usual unpolarized and polarized parton distributions, while the information encoded in $E$ and $\tilde E$ cannot be accessed since they are multiplied by the momentum transfer $\Delta$. Only in exclusive processes with a finite momentum transfer can $E$ and $\tilde E$ be probed.

We will study the unpolarized and polarized GPDs as well as their quasi counterparts defined in terms of spacelike correlations, and we will compute the one-loop corrections. Based on the one-loop results, we then propose a factorization formula for quasi GPDs, and extract the matching factors relating them to the light cone GPDs. The matching for the GPD $H$ ($\tilde H$) turns out to be similar to that for the parton distribution, whereas the matching for $E$ ($\tilde E$) is trivial since, as we will show in this paper, the quasi and light cone definition yields the same result for $E$ ($\tilde E$) at one-loop and leading power accuracy. This implies that the light cone GPD $E$ ($\tilde E$) can be smoothly approached by the large momentum limit of its quasi counterpart; hence, its simulation on the lattice is relatively simple. As a related quantity, we also present the matching for the distribution amplitude of the pion.

The rest of this paper is organized as follows. In Section II, we present the definitions of quasi GPDs and our conventions. In Section III, the results of our one-loop calculation for the unpolarized and polarized GPDs are given. The factorization formula for the quasi GPDs is presented in Section IV, where the one-loop matching factors are given. We also present the one-loop matching condition for the pion distribution amplitude. Section V contains our conclusions.

\section{quasi GPDs and conventions}
The quasi GPDs are defined in full analogy to the light cone ones, and they can be extracted from the following matrix elements defined on a spacelike interval along the $z$ direction~\cite{Ji:2013dva}
\begin{align}\label{quasiGPD}
{\cal F}_q(x,\xi,t,p^z)&=\int\frac{dz}{4\pi}e^{-ix p^z z}\langle p''|\bar\psi(-\frac{z}{2})\gamma^z L(-\frac{z}{2},\frac{z}{2})\psi(\frac{z}{2})|p'\rangle\non\\
&=\frac{1}{2p^z}\big[{\cal H}(x,\xi,t,p^z)\bar u(p'')\gamma^z u(p')+{\cal E}(x,\xi,t, p^z)\bar u(p'')\frac{i\sigma^{z\nu}\Delta_\nu}{2m}u(p')\big],\non\\
\tilde {\cal F}_q(x,\xi,t,p^z)&=\int\frac{dz}{4\pi}e^{-ix p^z z}\langle p''|\bar\psi(-\frac{z}{2})\gamma^z \gamma^5 L(-\frac{z}{2},\frac{z}{2})\psi(\frac{z}{2})|p'\rangle\non\\
&=\frac{1}{2p^z}\big[\tilde {\cal H}(x,\xi,t,p^z)\bar u(p'')\gamma^z \gamma^5 u(p')+\tilde {\cal E}(x,\xi,t, p^z)\bar u(p'')\frac{\gamma^5\Delta^z}{2m}u(p')\big].
\end{align}
The gauge link $L$ points along the $z$ direction, and $\cal H$, $\cal E$, $\tilde {\cal H}$ and $\tilde {\cal E}$ may depend on $p^z$. We define
\beq\label{quasivar}
p'^\mu=p^\mu-\frac{\Delta^\mu}{2}, \hspace{2em} p''^\mu=p^\mu+\frac{\Delta^\mu}{2}, \hspace{2em} p^\mu=(p^0,0,0,p^z), \hspace{2em} \xi=\frac{p''^z-p'^z}{p''^z+p'^z}=\frac{\Delta^z}{2p^z},
\eeq
and $t$ is the same as in the light cone GPDs, since it is Lorentz invariant. $\xi$ defined here approaches $\xi$ in the light cone GPDs when hadron's longitudinal momentum approaches infinity.

In the following we will focus on the generalized quark distributions in the nonsinglet case and consider quarks as external states. The on-shell conditions for the initial and final state quark
\beq
\big(p\pm\frac{\Delta}{2}\big)^2=m^2\non
\eeq
and the definition $t=\Delta^2$ lead to
\begin{align}
p^{0}= & \sqrt{m^{2}+p_{z}^{2}-\frac{t}{4}},\non\\
\Delta^{0}= & \frac{2\xi p_{z}^{2}}{\sqrt{m^{2}+p_{z}^{2}-\frac{t}{4}}},\non\\
\Delta^{1}= & \frac{\sqrt{-4t\left(1-\xi^{2}\right)p_{z}^{2}+t^{2}-4m^{2}\left(4\xi^{2}p_{z}^{2}+t\right)}}{\sqrt{4\left(m^{2}+p_{z}^{2}\right)-t}},
\end{align}
where we have kept a quark mass $m$ to regularize potential collinear singularities and have chosen $\vec\Delta_\perp$ to point in the positive $x$ direction without loss of generality. For $|\vec\Delta_\perp|$ to be real, we have the following constraint
\begin{align}
-4t\left(1-\xi^{2}\right)p_{z}^{2}+t^{2}-4m^{2}\left(4\xi^{2}p_{z}^{2}+t\right) & >0\implies\xi<\frac{1}{2p^{z}}\sqrt{\frac{-t\left(p_{z}^{2}+m^{2}-\frac{t}{4}\right)}{m^{2}-\frac{t}{4}}}.
\end{align}
In the infinite momentum limit $p^{z}\to\infty$, this reduces to
\beq
\xi <\sqrt{\frac{-t}{-t+4m^{2}}},
\eeq
which is the constraint for $\xi$ in the light cone GPDs. We will also assume $\xi>0$.

\section{one-loop result for GPDs}
In this section, we present the one-loop results for the quasi and light cone GPDs. As in the case of parton distributions, we choose the axial gauge $A^z=0$ throughout the computation since, in this gauge, the gauge link becomes unity. We also use a transverse momentum cutoff for regularizing the UV divergences.

Let us start with the unpolarized case. From the definition of Eq.~(\ref{quasiGPD}), it is easy to see that the quasi distributions yield the same result as the light cone ones at tree level
\begin{align}
H^{(0)}\left(x,\xi,t\right)={\cal H}^{(0)}\left(x,\xi,t, p^z\right)= & \delta\left(1-x\right),\non\\
E^{(0)}\left(x,\xi,t\right)={\cal E}^{(0)}\left(x,\xi,t, p^z\right)= & 0.
\end{align}

\begin{figure}[tbp]
\centering
\includegraphics[scale=0.4]{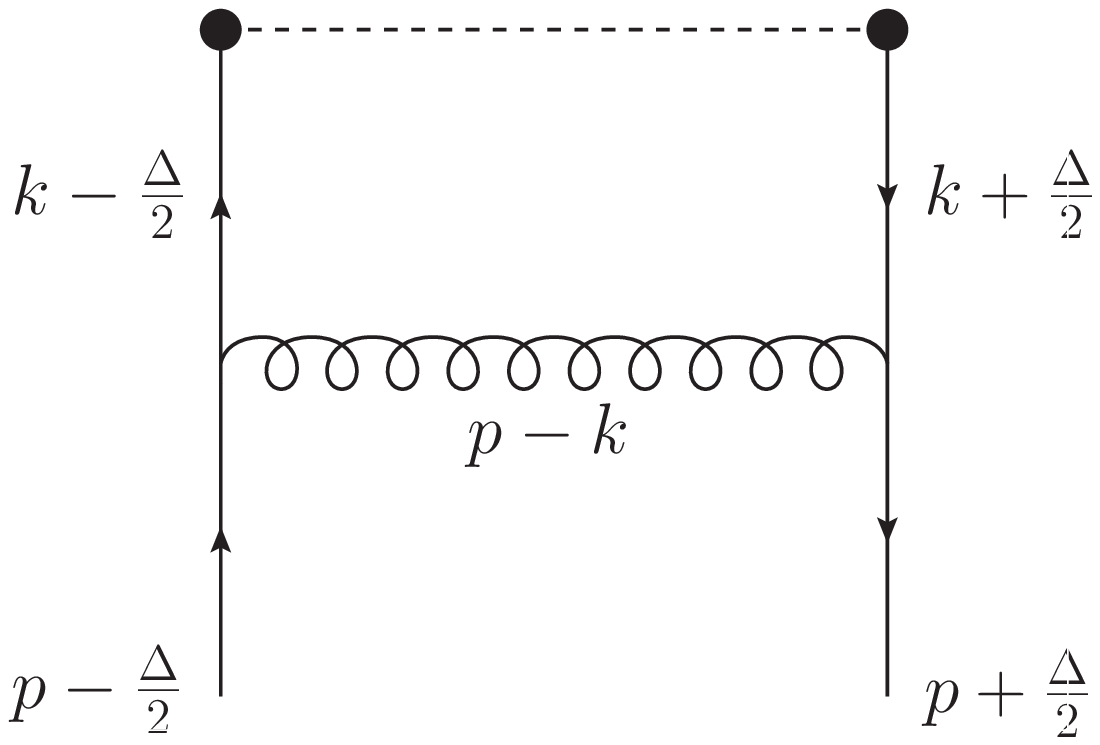}
\hspace{.2em}
\includegraphics[scale=0.6]{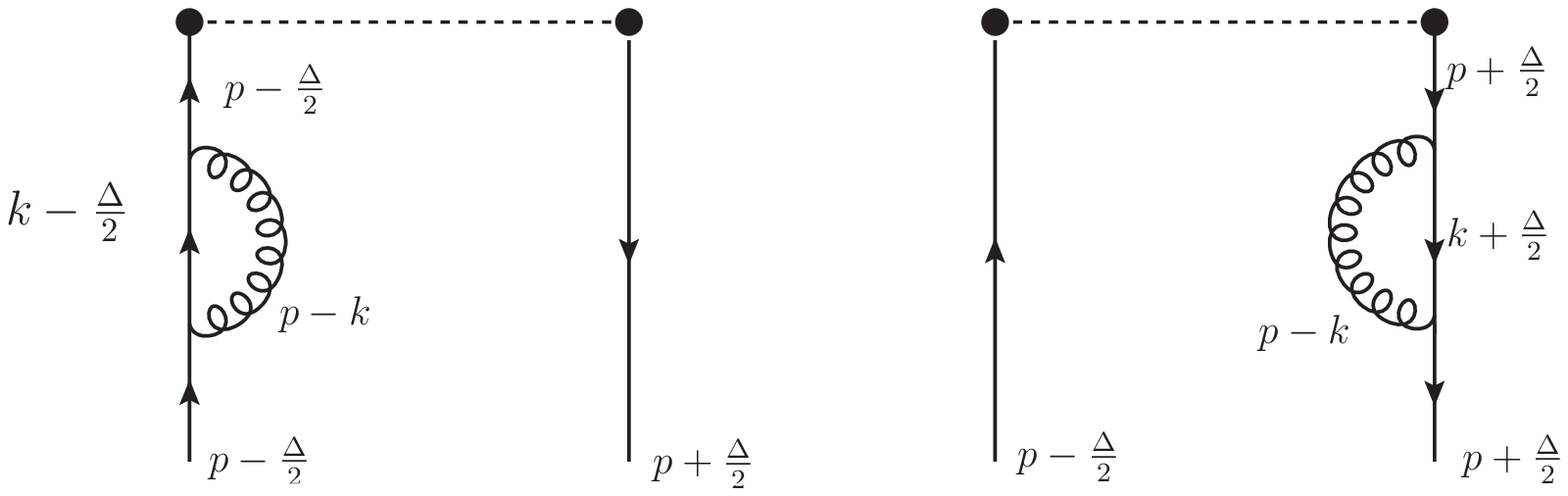}
\caption{One-loop diagrams for GPDs in the axial gauge.}
\label{1loopGPD}
\end{figure}

At one-loop level, the contributing Feynman diagrams in the axial gauge are shown in Fig.~\ref{1loopGPD}. Let us first look at the gluon-exchange diagram. In the axial gauge, the gluon propagator is given by $-i D_{\mu\nu}(k)/k^2$ with the numerator
\begin{align}\label{axialnum}
D_{\mu\nu}(k)=g_{\mu\nu}-\frac{n_\mu k_\nu+n_\nu k_\mu}{n\cdot k}+n^2\frac{k_\mu k_\nu}{(n\cdot k)^2},
\end{align}
where $n\cdot k=k^z, n^2=-1$. The first term on the rhs. of the above equation leads to the Feynman gauge result for the diagram, which can be written as
\begin{align}\label{fp}
\Gamma_1 &= C_F\int \frac{d^4k}{(2\pi)^4}\,\bar u(p'')\left(-ig\gamma^{\nu}\right)\frac{i}{\slashed k+\tfrac{\slashed\Delta}{2}-m}\gamma^{z}\frac{i}{\slashed k-\tfrac{\slashed\Delta}{2}-m}\left(-ig\gamma^{\mu}\right)\frac{-ig_{\mu\nu}}{\left(p-k\right)^{2}}u(p')\delta\big(x-\frac{k^z}{p^z}\big)\non\\
&=-ig^2C_F \int\frac{d^4k}{(2\pi)^4}\bar u(p'')\Big\{\frac{2\gamma^z}{[(k+\frac{\Delta}{2})^2-m^2](p-k)^2}\non\\
&+\frac{8m k^z-2\slashed k \gamma^z \slashed\Delta+2\slashed k \Delta^z-4k^z\slashed k-\gamma^z \Delta^2}{[(k+\frac{\Delta}{2})^2-m^2][(k-\frac{\Delta}{2})^2-m^2](p-k)^2}\Big\}u(p')\delta\big(x-\frac{k^z}{p^z}\big).
\end{align}

After a Feynman parametrization and integration over $k^0$ and $\vec k_\perp$, we have the following result for the first term in the curly brackets above
\begin{align}\label{gamma11}
\Gamma_{11}=\frac{g^2 C_F}{8\pi^2}p^z \int_0^1 dy\frac{\gamma^z}{\sqrt{((x-y)p^z+(1-y)\frac{\Delta^z}{2})^2+(1-y)^2 m^2}},
\end{align}
where we have used $k^z=x p^z$, and $y$ is the Feynman parameter.

The contribution of the second term in the curly brackets in Eq.~(\ref{fp}) can be computed analogously, and the result is
\begin{align}\label{gamma12}
\Gamma_{12}&=-\frac{g^2 C_F}{16\pi^2}\int dx\, p^z\int_0^1 dy\int_0^{1-y}dz \frac{1}{[(k^z-(1-z-y)p^z+(z-y)\frac{\Delta^z}{2})^2+4y z\, p^2+(z-y)^2 m^2]^{\frac{3}{2}}}\non\\
&\times\Big\{4m k^z-2(1-z-y)m(p^z+k^z-m\gamma^z)-(1-y)\gamma^z\Delta^2\non\\
&-(2k^z-\Delta^z)[(1-z-y)p^z\gamma^z-(z-y)\frac{\Delta^z}{2}\gamma^z-k^z\gamma^z]\Big\}\delta\big(x-\frac{k^z}{p^z}\big).
\end{align}
Performing the Feynman parameter integrations in Eqs.~(\ref{gamma11}) and (\ref{gamma12}), we are able to extract the contribution of $\Gamma_1$ to the quasi GPDs $\cal H$ and $\cal E$ with the help of the Gordon identity. The result reads
\begin{align}\label{Feyn}
{\cal H}_1\left(x,\xi,t,\mu,p^z\right) & =\non\\
&\hspace{-5em}\frac{\alpha_S C_F}{2\pi}\begin{cases}
\frac{\left(\xi^{2}+x\right)\ln\left(\frac{\xi+x}{x-\xi}\right)-\xi(2\xi-x+1)\ln\left(\frac{x-1}{x-\xi}\right)+\xi(2\xi+x-1)\ln\left(\frac{x-1}{\xi+x}\right)}{2\xi\left(\xi^{2}-1\right)} & \; x<-\xi\\
\\
-\frac{2\ln(m)(\xi+x)}{(\xi+1)(x-1)}+\frac{\ln(p^{z})(\xi+x)}{\xi(\xi+1)}-\frac{\ln(-t)(-2\xi+x-1)(\xi+x)}{2\xi(\xi+1)(x-1)}+\frac{(x-1)\ln(\xi+1)}{\xi^{2}-1}\\
+\frac{\ln(2)\left(\xi\left(x^{2}+1\right)-2x(\xi+x-1)\right)}{\xi\left(\xi^{2}-1\right)(x-1)}+\frac{\xi x\ln\left(1-\frac{x}{\xi}\right)}{\left(\xi^{2}-1\right)(x-1)}+\frac{(1-x)x\ln(\xi)}{\xi\left(\xi^{2}-1\right)(x-1)}+\frac{2\xi\ln\left(\frac{\xi+1}{1-x}\right)}{\xi^{2}-1}\\
+\frac{\xi\ln(\xi+x)}{\xi^{2}-1}+\frac{\xi\left(\xi^{2}-x\right)\ln(\frac{x+\xi}{\xi-x})-\xi^{2}\ln(\frac{\xi-x}{\xi})}{\xi\left(\xi^{2}-1\right)(x-1)} & \;-\xi<x<\xi\\
\\
\frac{4\ln(m)\left(x-\xi^{2}\right)}{(\xi-1)(\xi+1)(x-1)}+\frac{2(x-1)\ln(p^{z})}{\xi^{2}-1}-\frac{\ln(-t)\left(-2\xi^{2}+x^{2}+1\right)}{\left(\xi^{2}-1\right)(x-1)}+\frac{\ln(1-\xi)(-2\xi+x-1)}{\xi^{2}-1}\\
+\frac{\ln(\xi+1)(\xi+x-2)(\xi+x)}{\left(\xi^{2}-1\right)(x-1)}+\frac{1}{2}\left(\frac{2}{\xi+1}+\frac{(x-\xi)(\xi+x)}{\left(\xi^{2}-1\right)(x-1)}\right)\ln\left(-\frac{4(\xi+x)}{x-1}\right)\\
+\frac{(-2\xi+x-1)\ln\left(-\frac{4(x-\xi)}{x-1}\right)}{2\left(\xi^{2}-1\right)}+\frac{\left(\xi^{2}+x\right)\ln\left(\frac{2x}{\xi+x}-1\right)}{2\xi\left(\xi^{2}-1\right)}+\frac{\ln\left(\frac{1-x}{\xi+1}\right)}{x-1}+\frac{\ln\left(-\frac{4(\xi+x)}{(x-1)^{3}}\right)}{2(x-1)} & \;\xi<x<1\\
\\
-\frac{\left(\xi^{2}+x\right)\ln\left(\frac{\xi+x}{x-\xi}\right)-\xi(2\xi-x+1)\ln\left(\frac{x-1}{x-\xi}\right)+\xi(2\xi+x-1)\ln\left(\frac{x-1}{\xi+x}\right)}{2\xi\left(\xi^{2}-1\right)} & x>1,
\end{cases}
\end{align}
\begin{align}\label{E1}
{\cal E}_1\left(x,\xi,t,\mu,p^z\right) & =\non\\
&\hspace{-5em} \frac{\alpha_S C_F}{2\pi}\frac{m^{2}}{\left(1-\xi^{2}\right)t}\begin{cases}
0 & x<-\xi\\
2\left(\xi-1\right)\left(x-\xi\right)\ln\left(\frac{-t}{m^{2}}\right)+2\left(\xi^{2}+x\right)\ln \frac{\xi-x}{\xi+x}\\
+2\xi\left(x+1\right)\ln\frac{\left(1+\xi\right)^{2}\left(\xi^{2}-x^{2}\right)}{4\xi^{2}\left(1-x\right)^{2}} & -\xi<x<\xi\\
4\left(\xi^{2}+x\right)\ln\left(\frac{m^{2}}{-t}\right)+4\xi\left(1+x\right)\ln\frac{1+\xi}{1-\xi} & \xi<x<1\\
0 & x>1,
\end{cases}
\end{align}
where $\mu$ is the transverse momentum cutoff for regularizing potential UV divergences. Some remarks on the above results are in order. To obtain Eqs.~(\ref{Feyn}) and (\ref{E1}), we take the limit $\mu\to\infty$, and then $p^z\to\infty$, $m\to 0$, where we keep the leading term in $\mu$, $p^z$ and $m$ of ${\cal H}_1$ and ${\cal E}_1$ (for ${\cal E}_1$ the leading term is $\mathcal O(m^2)$), and ignore the power suppressed corrections of the type $(1/p^z)^n (n\ge1)$. If we do not take $p^z\to\infty$, power corrections should be kept in the light cone GPDs as well, in order for the quasi and light cone GPDs to have the same IR behavior. As in the PDF case~\cite{Xiong:2013bka, Lin:2014zya}, the quasi GPD result ${\cal H}_1$ does not vanish in the full $x$ range. However, its collinear singularities exist only in the Dokshitzer-Gribov-Lipatov-Altarelli-Paris (DGLAP) and Efremov-Radyushkin-Brodsky-Lepage (ERBL) regions $\xi<x<1$ and $-\xi<x<\xi$. In the DGLAP region, the collinear singularities do not straightforwardly reduce to the corresponding PDF result with the Altarelli-Parisi kernel when taking the limit $\xi, t\to 0$, because we assume a finite $t$ in taking the limit $m\to 0$. If $t=0$, as in the PDF case, the above assumption does not apply, and the term leading to $\ln(-t)$ above will lead to $\ln m^2$, and thus to the correct collinear behavior of PDF. $\cal E$ is UV convergent as expected, since it is zero at tree level. There is no UV divergence in $\cal H$, but a logarithmic dependence on $p^z$ instead, as in the one-loop results for PDFs. Moreover, the coefficient of $\ln p^z$ in $\cal H$ in the DGLAP region reduces to the corresponding PDF result when $\xi=0$.

The second numerator structure in Eq.~(\ref{axialnum}) leads to
\begin{align}\label{sp}
\Gamma_2&=C_F \int \frac{d^{4}k}{(2\pi)^4}\,\bar u(p'')\left(-ig\gamma^{\nu}\right)\frac{i}{\slashed k+\tfrac{\slashed\Delta}{2}-m}\gamma^{z}\frac{i}{\slashed k-\tfrac{\slashed\Delta}{2}-m}\left(-ig\gamma^{\mu}\right)\non\\
&\times\frac{i\left[\left(p-k\right)_{\mu}n_{\nu}+n_{\mu}\left(p-k\right)_{\nu}\right]}{\left(p^{z}-k^{z}\right)\left(p-k\right)^{2}}u(p')\delta\big(x-\frac{k^z}{p^z}\big)\non\\
&=-i g^2 C_F \int \frac{d^{4}k}{(2\pi)^4}\,\bar u(p'')\left[\gamma^{z}\frac{1}{\slashed k+\tfrac{\slashed\Delta}{2}-m}\gamma^{z}+\gamma^{z}\frac{1}{\slashed k-\tfrac{\slashed\Delta}{2}-m}\gamma^{z}\right]\frac{1}{\left(p^{z}-k^{z}\right)\left(p-k\right)^{2}}u(p')\delta\big(x-\frac{k^z}{p^z}\big),
\end{align}
and the result is
\begin{align}
{\cal H}_2\left(x,\xi,t,\mu,p^z\right)& =\non\\
&\hspace{-5em}  \frac{\alpha_S C_{F}}{2\pi}\begin{cases}
\frac{1}{1-x}+\frac{x-\xi}{\left(1-x\right)\left(1-\xi\right)}\ln\frac{x-1}{x-\xi}+\frac{x+\xi}{\left(1-x\right)\left(1+\xi\right)}\ln\frac{x-1}{x+\xi} & x<-\xi\\
\\
\frac{x+\xi}{\left(1-x\right)\left(1+\xi\right)}\left( \ln\frac{p_{z}^{2}}{m^{2}}+\ln\frac{4\left(x+\xi\right)\left(1+\xi\right)^{2}}{1-x}-\frac{1}{2}\right)+\frac{1}{2\left(1+\xi\right)}\non\\
+\frac{1}{2\left(1-x\right)}+\frac{x-\xi}{\left(1-x\right)\left(1-\xi\right)}\ln\frac{x-1}{x-\xi} & -\xi<x<\xi\\
\\
\frac{2\left(x-\xi^{2}\right)}{\left(1-x\right)\left(1-\xi^{2}\right)}\left(\ln\frac{p_{z}^{2}}{m^{2}}-\frac{1}{2}\right)+\frac{\left(x-\xi\right)}{\left(1-x\right)\left(1-\xi\right)}\ln\frac{4\left(x-\xi\right)\left(1-\xi\right)^{2}}{1-x}\\
+\frac{\left(x+\xi\right)}{\left(1-x\right)\left(1+\xi\right)}\ln\frac{4\left(x+\xi\right)\left(1+\xi\right)^{2}}{1-x}+\frac{1}{1-\xi^{2}} & \xi<x<1\\
-\frac{1}{1-x}-\frac{x-\xi}{\left(1-x\right)\left(1-\xi\right)}\ln\frac{x-1}{x-\xi}-\frac{x+\xi}{\left(1-x\right)\left(1+\xi\right)}\ln\frac{x-1}{x+\xi} & x>1,
\end{cases}\non\\
{\cal E}_2(x,\xi,t,\mu,p^z)&=0,
\end{align}
where ${\cal E}_2(x,\xi,t,\mu,p^z)$ is of order ${\mathcal O}(m^2/p_z^2)$, which is power suppressed compared to ${\cal E}_1$ in Eq.~(\ref{E1}) and therefore ignored.

The last numerator structure yields
\begin{align}\label{dp}
\Gamma_3&=C_F \int \frac{d^{4}k}{(2\pi)^4}\bar u(p'')\left(-ig\gamma^{\nu}\right)\frac{i}{\slashed k+\tfrac{\slashed\Delta}{2}-m}\gamma^{z}\frac{i}{\slashed k-\tfrac{\slashed\Delta}{2}-m}\left(-ig\gamma^{\mu}\right)\frac{i\left(p-k\right)_{\mu}\left(p-k\right)_{\nu}}{\left(p-k\right)^{2}\left(p^{z}-k^{z}\right)^{2}}u(p')\delta\big(x-\frac{k^z}{p^z}\big)		\non\\
&=ig^2 C_F \int \frac{d^{4}k}{(2\pi)^4}\bar u(p'')\frac{\gamma^{z}}{\left(p-k\right)^{2}\left(p^{z}-k^{z}\right)^{2}}u(p')\delta\big(x-\frac{k^z}{p^z}\big),
\end{align}
which contributes to ${\cal H}\left(x,\xi,t,\mu,p^z\right)$ only with
\begin{align}
{\cal H}_3(x,\xi,t,\mu,p^z)= & \frac{\alpha_{S}C_F}{2\pi}\frac{\sqrt{\mu^{2}+p_{z}^{2}\left(1-x\right)^{2}}-\left|1-x\right|p^{z}}{p^{z}\left(1-x\right)^{2}}.
\end{align}

Summing over all of these contributions, we obtain the following results for the gluon-exchange diagram in Fig.~\ref{1loopGPD}:
\begin{align}
{\cal H}^{(1)}(x,\xi,t,\mu,p^z)& =\non\\
&\hspace{-5em}  \frac{\alpha_S C_F}{2\pi}\begin{cases}
\frac{(\xi^{2}+x)\ln\frac{\xi+x}{x-\xi}}{2\xi(\xi^{2}-1)}+\frac{(-2\xi^{2}+x^{2}+1)\ln\frac{(x-1)^{2}}{x^{2}-\xi^{2}}}{2\left(\xi^{2}-1\right)(x-1)}+\frac{\mu}{p^z(1-x)^{2}} & x<-\xi \\
\frac{x+\xi}{2\xi(1+\xi)}(1+\frac{2\xi}{1-x})\ln\frac{p_{z}^{2}}{-t}+\frac{1+x^2-2\xi^2}{(1-x)(1-\xi^2)}\ln[{2(\xi+1)}]-\frac{x+\xi^2}{\xi(\xi^2-1)}\ln{4\xi}\non\\
+\frac{x+\xi}{(1+\xi)(x-1)}+\frac{\mu}{p^z(1-x)^{2}} &-\xi<x<\xi \\
\frac{1+x^2-2\xi^2}{(1-x)(1-\xi^2)}\ln\frac{p_z^2}{-t}+\frac{1+x^2-2\xi^2}{2(1-x)(1-\xi^2)}\big(\ln[16(x^2-\xi^2)]-2\ln\frac{1-x}{1-\xi^2}\big)\non\\
-\frac{x+\xi^2}{2\xi(1-\xi^2)}\ln\frac{x-\xi}{x+\xi}-\frac{2(x-\xi^2)}{(1-x)(1-\xi^2)}+\frac{\mu}{p^z(1-x)^{2}} & \xi<x<1 \\
-\frac{(\xi^{2}+x)\ln\frac{\xi+x}{x-\xi}}{2\xi(\xi^{2}-1)}-\frac{(-2\xi^{2}+x^{2}+1)\ln\frac{(x-1)^{2}}{x^{2}-\xi^{2}}}{2\left(\xi^{2}-1\right)(x-1)}+\frac{\mu}{p^z(1-x)^{2}} & x>1,
\end{cases}\non\\
{\cal E}^{(1)}(x,\xi,t,\mu,p^z)&={\cal E}_1(x,\xi,t,\mu,p^z).
\end{align}

Now we present the one-loop results for light cone GPDs. As in the PDF case~\cite{Xiong:2013bka}, the one-loop corrections for light cone GPDs can be obtained by first integrating over $k^0$, then taking the limit $p^z\to\infty$ and integrating over $\vec k_\perp$. This leads to the following results for the three numerator structures in Eq.~(\ref{axialnum})
\begin{align}
H_1 \left(x,\xi,t,\mu\right)& =\non\\
&\hspace{-5em} \frac{\alpha_S C_F}{2\pi}\begin{cases}
\frac{x+\xi}{2\xi\left(1+\xi\right)}\ln\left(\frac{\mu^{2}}{-t}\right)+\frac{x+\xi}{\left(1-x\right)\left(1+\xi\right)}\ln\left(\frac{m^{2}}{-t}\right)-\frac{1-x-2\xi}{1-\xi^{2}}\ln\frac{1-x}{1+\xi}\\
+\frac{x+\xi^{2}}{2\xi\left(1-\xi^{2}\right)}\ln\frac{4\xi^{2}}{\xi^{2}-x^{2}}+\frac{1+x^{2}-2\xi^{2}}{2\left(1-x\right)\left(1-\xi^{2}\right)}\ln\frac{\xi-x}{x+\xi} & -\xi<x<\xi\\
\frac{1-x}{\left(1-\xi^{2}\right)}\ln\left(\frac{\mu^{2}}{-t}\right)+\frac{2\left(x-\xi^{2}\right)}{\left(1-x\right)\left(1-\xi^{2}\right)}\ln\left(\frac{m^{2}}{-t}\right) & \;\\
-\frac{1-x-2\xi}{1-\xi^{2}}\ln\frac{1-x}{1+\xi}-\frac{1-x+2\xi}{1-\xi^{2}}\ln\frac{1-x}{1-\xi} & \;\xi<x<1\\
0 & \;\text{otherwise},
\end{cases}\non\\
E_1\left(x,\xi,t,\mu\right)&= \frac{\alpha_{S}C_F}{2\pi}\frac{m^{2}}{-t}\begin{cases}
\frac{2\left(x-\xi\right)}{1+\xi}\ln\left(\frac{-t}{m^{2}}\right)+\frac{2\xi\left(1+x\right)}{1-\xi^{2}}\ln\frac{4\xi^{2}\left(1-x\right)^{2}}{\left(1+\xi\right)^{2}\left(\xi^{2}-x^{2}\right)}\\
+\frac{2\left(x+\xi^{2}\right)}{1-\xi^{2}}\ln\frac{x+\xi}{\xi-x} & -\xi<x<\xi\\
\frac{4\left(x+\xi^{2}\right)}{1-\xi^{2}}\ln\left(\frac{-t}{m^{2}}\right)+\frac{4\xi\left(1+x\right)}{1-\xi^{2}}\ln\frac{1-\xi}{1+\xi} & \;\xi<x<1\\
0 & \;\text{otherwise},
\end{cases}\non
\end{align}
\begin{align}
H_2\left(x,\xi,t,\mu\right)& =\non\\
&\hspace{-5em} \frac{\alpha_{S}C_F}{2\pi}\begin{cases}
\frac{x+\xi}{\left(1-x\right)\left(1+\xi\right)}\left\{ \ln\frac{\mu^{2}}{m^{2}}+\ln\left[\left(\frac{1+\xi}{1-x}\right)^{2}\right]\right\} \;\;\; & -\xi<x<\xi\\
\frac{2\left(x-\xi^{2}\right)}{\left(1-x\right)\left(1-\xi^{2}\right)}\ln\frac{\mu^{2}}{m^{2}}+\frac{x+\xi}{\left(1-x\right)\left(1+\xi\right)}\ln\left[\left(\frac{1+\xi}{1-x}\right)^{2}\right]\\
+\frac{x-\xi}{\left(1-x\right)\left(1-\xi\right)}\ln\left[\left(\frac{1-\xi}{1-x}\right)^{2}\right] & \xi<x<1\\
0 & \text{otherwise},
\end{cases}\non\\
E_2\left(x,\xi,t,\mu\right)&= 0,\non\\
H_3\left(x,\xi,t,\mu\right)&= 0,
\end{align}
the sum of which is
\begin{align}
H^{(1)}(x,\xi,t,\mu)&= \frac{\alpha_S C_F}{2\pi}\begin{cases}
\frac{x+\xi}{2\xi(1+\xi)}(1+\frac{2\xi}{1-x})\ln\frac{\mu^{2}}{-t}+\frac{x+\xi^2}{2\xi(1-\xi^2)}\ln\frac{4\xi^2}{\xi^2-x^2}\non\\
-\frac{1+x^2-2\xi^2}{2(1-x)(1-\xi^2)}(\ln\frac{(1-x)^2}{(1+\xi)^2}-\ln\frac{\xi-x}{x+\xi}) &-\xi<x<\xi \\
\frac{1+x^2-2\xi^2}{(1-x)(1-\xi^2)}\ln\frac{\mu^2}{-t}-\frac{1+x^2-2\xi^2}{(1-x)(1-\xi^2)}\ln\frac{(1-x)^2}{(1-\xi^2)} & \xi<x<1 \\
0 & \text{otherwise},
\end{cases}\non\\
E^{(1)}(x,\xi,t,\mu)&=E_1(x,\xi,t,\mu).
\end{align}
The results do not vanish only in the DGLAP and ERBL regions. There is a logarithmic UV divergence $\ln\mu^2$ in the above results, whose coefficient agrees with the coefficient of $\ln p_z^2$ in ${\cal H}^{(1)}$, and also with the evolution kernel of nonsinglet GPDs in the DGLAP and ERBL regions. It is interesting to see that $E^{(1)}$ and ${\cal E}^{(1)}$ are equal. 
This means no matching is required for the GPD $E$ up to one-loop and leading power accuracy; therefore, the light cone $E$ can be smoothly approached by the quasi $\cal E$ in the large momentum limit. The reason behind this is simple: the light cone GPD $E$ and its quasi counterpart are zero at tree level and thus they are UV convergent at one-loop level.	

Next, we look at the contribution of self-energy diagrams in Fig.~\ref{1loopGPD}. The computation of the quark wave function renormalization factor is essentially the same as in the PDF case~\cite{Xiong:2013bka}; the only difference is in the momenta of the quarks. Since the incoming and outgoing quarks now have different momenta, we have two wave function renormalization factors $\frac1 2\delta {\cal Z}_{F}\left(p\pm\frac{1}{2}\Delta\right)$ (or $\frac1 2\delta Z_{F}\left(p\pm\frac{1}{2}\Delta\right)$ for light cone GPDs). For simplicity, we denote $\delta {\cal Z}_{F}\left(p\pm\frac{1}{2}\Delta\right)$ ($\delta Z_{F}\left(p\pm\frac{1}{2}\Delta\right)$) as $\delta {\cal Z}_{F,\pm}\left(\xi,t\right)$ ($\delta Z_{F,\pm}\left(\xi,t\right)$). These factors can be related to each other as
\beq
\delta {\cal Z}_{F,+}\left(\xi,t\right)= \delta {\cal Z}_{F,-}\left(-\xi,t\right), \hspace{2em} \delta  Z_{F,+}\left(\xi,t\right)= \delta  Z_{F,-}\left(-\xi,t\right).
\eeq
Using the same strategy as in Ref.~\cite{Xiong:2013bka}, we obtain the following results for the wave function renormalization factors
\begin{align}
\delta{\cal Z}_{F1,-}= & -\frac{\alpha_{S}C_F}{2\pi}\int dx\,\begin{cases}
\frac{x-1}{\left(1-\xi\right)^{2}}\ln\frac{x-\xi}{x-1}-\frac{2}{1-\xi} & x<\xi\\
\frac{1-x}{\left(1-\xi\right)^{2}}\ln\frac{p_{z}^{2}}{m^{2}}+\frac{1-x}{\left(1-\xi\right)^{2}}\ln\frac{4\left(1-\xi\right)^{2}\left(x-\xi\right)}{1-x}\;\;\;\\
-\frac{2\left(x-\xi\right)^{2}}{\left(1-x\right)\left(1-\xi\right)^{2}}-\frac{2\left(1-x\right)}{\left(1-\xi\right)^{2}} & \xi<x<1\\
-\frac{x-1}{\left(1-\xi\right)^{2}}\ln\frac{x-\xi}{x-1}+\frac{2}{1-\xi} & x>1,
\end{cases}\non\\
\delta{\cal Z}_{F2,-}= & -\frac{\alpha_{S}C_F}{2\pi}\int dx\,\begin{cases}
\frac{2\left(x-\xi\right)}{\left(x-1\right)\left(1-\xi\right)}\ln\frac{x-\xi}{x-1}+\frac{3-2x-\xi}{\left(x-1\right)\left(\xi-1\right)} & x<\xi\\
\frac{2\left(x-\xi\right)}{\left(1-x\right)\left(1-\xi\right)}\left( \ln\frac{p_z^2}{m^2}+\ln\frac{4\left(1-\xi\right)^{2}\left(x-\xi\right)}{1-x}\right) \;\;\; & \xi<x<1\\
+\frac{\xi-4x+3}{\left(x-1\right)\left(\xi-1\right)}\\
-\frac{2\left(x-\xi\right)}{\left(x-1\right)\left(1-\xi\right)}\ln\frac{x-\xi}{x-1}-\frac{3-2x-\xi}{\left(x-1\right)\left(\xi-1\right)} & x>1,
\end{cases}\non
\end{align}
\begin{align}
\delta{\cal Z}_{F3,-}= & -\frac{\alpha_{S}C_F}{2\pi}\int dx\,\begin{cases}
\frac{\mu}{\left(1-x\right)^{2}p^{z}}+\frac{x+\xi-2}{\left(x-1\right)\left(\xi-1\right)}\;\;\; & x<\xi\\
\frac{\mu}{\left(1-x\right)^{2}p^{z}}+\frac{x+\xi-2}{\left(x-1\right)\left(\xi-1\right)} & \xi<x<1\\
\frac{\mu}{\left(1-x\right)^{2}p^{z}}-\frac{x+\xi-2}{\left(x-1\right)\left(\xi-1\right)} & x>1.
\end{cases}
\end{align}
Summing over all these contributions and including also $\delta {\cal Z}_{F,+}$, we have
\begin{align}
{\cal Z}_F^{(1)}&= -\frac{\alpha_S C_F}{2\pi}\int dy\begin{cases}
\big(f(\xi,y)\ln\frac{y-\xi}{y-1}+f(-\xi,y)\ln\frac{y+\xi}{y-1}\big)-\frac{1}{1-\xi^2}+\frac{\mu}{p^z(1-y)^{2}} & y<-\xi \\
-f(-\xi,y)\ln\frac{p_z^2}{m^2}-f(\xi,y)\ln\frac{1-y}{\xi-y}+f(-\xi, y)\ln\frac{1-y}{4(1+\xi)^2(\xi+y)}\non\\
+4f(-\xi, y)-\frac{1}{1-\xi^2}+\frac{2}{1-y}+\frac{\mu}{p^z(1-y)^{2}} &-\xi<y<\xi \\
-(f(\xi, y)+f(-\xi, y))\ln\frac{p_z^2}{m^2}+f(\xi, y)\ln\frac{1-y}{4(y-\xi)(1-\xi)^2}\non\\
+f(-\xi, y)\ln\frac{1-y}{4(y+\xi)(1+\xi)^2}+4(f(\xi, y)+f(-\xi, y))+\frac{4}{1-y}-\frac{1}{1-\xi^2}\non\\
+\frac{\mu}{p^z(1-y)^{2}} & \xi<y<1 \\
-f(\xi, y)\ln\frac{y-\xi}{y-1}-f(-\xi, y)\ln\frac{y+\xi}{y-1}+\frac{1}{1-\xi^2}+\frac{\mu}{p^z(1-y)^{2}} & y>1,
\end{cases}
\end{align}
where
\begin{align}
f(\xi, y)&=\frac{1}{1-\xi}-\frac{1}{1-y}-\frac{1-y}{2(1-\xi)^2}.
\end{align}

Similarly, on the light cone one obtains
\begin{align}
\delta Z_{F1,-}= & -\frac{\alpha_{S}C_F}{2\pi}\int dx\,\begin{cases}
\frac{1-x}{\left(1-\xi\right)^{2}}\left(\ln\frac{\mu^{2}}{m^{2}}+2\ln\frac{1-\xi}{1-x}\right)-\frac{2\left(x-\xi\right)}{\left(1-x\right)\left(1-\xi\right)}\;\;\; & \xi<x<1\\
0 & \text{otherwise},
\end{cases}\non\\
\delta Z_{F2,-}= & -\frac{\alpha_{S}C_F}{2\pi}\int dx\,\begin{cases}
\frac{2\left(x-\xi\right)}{\left(1-\xi\right)\left(1-x\right)}\left(\ln\frac{\mu^{2}}{m^{2}}+2\ln\frac{1-\xi}{1-x}\right)\;\;\; & \xi<x<1\\
0 & \text{otherwise},
\end{cases}\non\\
\delta Z_{F3,-}= & 0,
\end{align}
and the complete result including $\delta Z_{F,+}$ is
\begin{align}
Z_F^{(1)}&= -\frac{\alpha_S C_F}{2\pi}\int dy\begin{cases}
-f(-\xi, y)\ln\frac{\mu^2}{m^2}-2f(-\xi, y)\ln\frac{1+\xi}{1-y}+\frac{1}{1+\xi}-\frac{1}{1-y} &-\xi<y<\xi \\
-(f(\xi, y)+f(-\xi, y))\ln\frac{\mu^2}{m^2}-2f(\xi, y)\ln\frac{1-\xi}{1-y}\non\\
-2f(-\xi, y)\ln\frac{1+\xi}{1-y}+\frac{1}{1-\xi}+\frac{1}{1+\xi}-\frac{2}{1-y} & \xi<y<1 \\
0 & \text{otherwise}.
\end{cases}
\end{align}
From the above results, it is clear that the coefficients of $\ln\frac{p_z^2}{m^2}$ in ${\cal Z}_F^{(1)}$ and $\ln\frac{\mu^2}{m^2}$ in $Z_F^{(1)}$ in the DGLAP region $[\xi, 1]$ reduce to the PDF result when $\xi\to 0$. Several further checks have been performed on our results: the one-loop results for GPDs reduce to that for PDF when $\xi, t\to 0$ (although for $H$, this is non-trivial); the $x$ integrals of the Feynman part Eqs.~(\ref{fp}) contribution to GPDs, which is the most complicated one among all three numerator structures, agree with the form factors of a local quark current computed in the Feynman gauge. We also checked the polynomiality of the logarithmic terms in the above results.

In the case of polarized GPDs, the vertex contribution can be obtained by replacing $\gamma^z$ with $\gamma^z \gamma^5$ in Eq.~(\ref{fp}), whereas the self-energy contribution remains the same. We present the result of the vertex contribution below. For the quasi GPDs, we have
\begin{align}
\tilde {\cal H}^{(1)}(x,\xi,t,\mu,p^z)& = {\cal H}^{(1)}(x,\xi,t,\mu,p^z),\non\\
\tilde {\cal E}^{(1)}(x,\xi,t,\mu,p^z)&=\non\\
&\hspace{-5em} \frac{\alpha_S C_F}{2\pi}\frac{m^{2}}{\xi\left(1-\xi^{2}\right)t}\begin{cases}
0 & x<-\xi\\
2\left(\xi-1\right)\left(\xi-x\right)\ln\left(\frac{-t}{m^{2}}\right)+2\xi\left(1+x\right)\ln\frac{\xi-x}{\xi+x}\\
+2\left(\xi^2+x\right)\ln\frac{\left(1+\xi\right)^{2}\left(\xi^{2}-x^{2}\right)}{4\xi^{2}\left(1-x\right)^{2}} & -\xi<x<\xi\\
4\xi\left(1+x\right)\ln\left(\frac{m^{2}}{-t}\right)+4\left(\xi^2+x\right)\ln\frac{1+\xi}{1-\xi} & \xi<x<1\\
0 & x>1,
\end{cases}
\end{align}
whereas on the light cone we have
\begin{align}
\tilde H^{(1)}(x,\xi,t,\mu,p^z)& = H^{(1)}(x,\xi,t,\mu,p^z),\non\\
\tilde E^{(1)}(x,\xi,t,\mu,p^z)&=\tilde {\cal E}^{(1)}(x,\xi,t,\mu,p^z).
\end{align}

\section{One-Loop Factorization}
Before we construct the factorization formula connecting the quasi and light cone GPDs, let us summarize the one-loop results in the previous section. For the unpolarized case, we have
\begin{align}
{\cal H}(x,\xi,t,\mu,p^z)&=[1+\frac{1}{2}(\delta {\cal Z}_F(p-\frac{1}{2}\Delta)+\delta {\cal Z}_F(p+\frac{1}{2}\Delta))]\delta(1-x)+{\cal H}^{(1)}(x,\xi,t,\mu,p^z)\non\\
&=(1+{\cal Z}_F^{(1)})\delta(1-x)+{\cal H}^{(1)}(x,\xi,t,\mu,p^z),\non\\
H(x,\xi,t,\mu)&=[1+\frac{1}{2}(\delta Z_F(p-\frac{1}{2}\Delta)+\delta Z_F(p+\frac{1}{2}\Delta))]\delta(1-x)+H^{(1)}(x,\xi,t,\mu)\non\\
&=(1+Z_F^{(1)})\delta(1-x)+H^{(1)}(x,\xi,t,\mu).
\end{align}
Similar results can be written down for the polarized case. A crucial difference between $\cal H$ and $H$ is that the latter vanishes in the regions $x<-\xi$ and $x>1$, whereas the former does not. The connection between the quasi and light cone GPDs can be established as
\begin{align}\label{oneloopfact}
{\cal H}(x,\xi,t,\mu,p^z)&=\int_{-1}^1 \frac{dy}{|y|}Z_H\left(\frac{x}{y},\frac{\xi}{y},\frac{\mu}{p^z}\right) H(y,\xi,t,\mu)
\end{align}
up to power corrections suppressed by $p^z$, where the integration range is given by the support property of the light cone GPD.

The matching factor $Z_H$ can be perturbatively expanded as
\beq
Z_H\left(\frac{x}{y}, \frac{\xi}{y},\frac{\mu}{p^z}\right) = \delta\left(1-\frac{x}{y}\right) + \frac{\alpha_S}{2\pi} Z_H^{(1)}\left(\frac{x}{y},\frac{\xi}{y},
\frac{\mu}{p^z}\right) + h.o.,
\eeq
where $h.o.$ denotes higher-order contributions. From the one-loop results for $\cal H$ and $H$, the matching factor can be extracted as
\begin{align}\label{1loopmatfac}
Z_H^{(1)}(\eta, \zeta, \mu/p^z)/C_F&=\begin{cases}
\frac{(\zeta^{2}+\eta)\ln\frac{\zeta+\eta}{\eta-\zeta}}{2\zeta(\zeta^{2}-1)}+\frac{(-2\zeta^{2}+\eta^{2}+1)\ln\frac{(\eta-1)^{2}}{\eta^{2}-\zeta^{2}}}{2\left(\zeta^{2}-1\right)(\eta-1)}+\frac{\mu}{p^z(1-\eta)^{2}} & \eta<-\zeta \\
\frac{\eta+\zeta}{2\zeta(1+\zeta)}(1+\frac{2\zeta}{1-\eta})\ln\frac{p_{z}^{2}}{\mu^2}+\frac{1+\eta^2-2\zeta^2}{2(1-\eta)(1-\zeta^2)}\big(\ln[4(1-\eta)^2]-\ln\frac{\zeta-\eta}{\eta+\zeta}\big)\non\\
+\frac{\eta+\zeta^2}{2\zeta(1-\zeta^2)}\ln{[4(\zeta^2-\eta^2)]}+\frac{\eta+\zeta}{(1+\zeta)(\eta-1)}+\frac{\mu}{p^z(1-\eta)^{2}} &-\zeta<\eta<\zeta \\
\frac{1+\eta^2-2\zeta^2}{(1-\eta)(1-\zeta^2)}\ln\frac{p_z^2}{\mu^2}+\frac{1+\eta^2-2\zeta^2}{2(1-\eta)(1-\zeta^2)}\big(\ln[16(\eta^2-\zeta^2)]+2\ln(1-\eta)\big)\non\\
-\frac{\eta+\zeta^2}{2\zeta(1-\zeta^2)}\ln\frac{\eta-\zeta}{\eta+\zeta}-\frac{2(\eta-\zeta^2)}{(1-\eta)(1-\zeta^2)}+\frac{\mu}{p^z(1-\eta)^{2}} & \zeta<\eta<1 \\
-\frac{(\zeta^{2}+\eta)\ln\frac{\zeta+\eta}{\eta-\zeta}}{2\zeta(\zeta^{2}-1)}-\frac{(-2\zeta^{2}+\eta^{2}+1)\ln\frac{(\eta-1)^{2}}{\eta^{2}-\zeta^{2}}}{2\left(\zeta^{2}-1\right)(\eta-1)}+\frac{\mu}{p^z(1-\eta)^{2}} & \eta>1.
\end{cases}\\
\end{align}
The above matching factor is valid for $y>\xi$, however, it can be extended to the full $y$ range as
\begin{align}\label{1loopmatchingfull}
\frac{1}{|y|}Z_H^{(1)}\left(\frac{x}{y}, \frac{\xi}{y},\frac{\mu}{p^z}\right)/C_F&=\frac{1}{y}\Big[F_1\left(\frac{x}{y}, \frac{\xi}{y},\frac{\mu}{p^z}\right)\theta(x<-\xi)\theta(x<y)\non\\
&+F_2\left(\frac{x}{y}, \frac{\xi}{y},\frac{\mu}{p^z}\right)\theta(-\xi<x<\xi)\theta(x<y)\non\\
&+F_3\left(\frac{x}{y}, \frac{\xi}{y},\frac{\mu}{p^z}\right)\theta(\xi<x<y)+F_4\left(\frac{x}{y}, \frac{\xi}{y},\frac{\mu}{p^z}\right)\theta(x>\xi)\theta(x>y)\Big],
\end{align}
where $F_{1,2,3,4}$ are given by the matching factor in the four different regions in Eq.~(\ref{1loopmatfac}) ( with a replacement $\ln a\to 1/2\ln a^2$ so that they are real functions); $p^z$ shall be replaced by $y P^z$ with $P^z$ being the averaged longitudinal momentum of the external hadrons. The validity of the above equation can be checked by explicit computations. The coefficient of $\ln\frac{p_z^2}{\mu^2}$ is the same as the evolution kernel of the light cone GPD.


Near $\eta=1$, one has an extra contribution from the self-energy correction
\beq
Z_H^{(1)}\left(\frac{x}{y}, \frac{\xi}{y},\frac{\mu}{p^z}\right)=\delta Z_H^{(1)}(2\pi/\alpha_S)\delta\left(1-\eta\right)
\eeq
with
\begin{align}\label{1loopmatfacwf}
\delta Z_H^{(1)}&=-\frac{\alpha_S C_F}{2\pi}\int d\eta\begin{cases}
\big(f(\zeta,\eta)\ln\frac{\eta-\zeta}{\eta-1}+f(-\zeta,\eta)\ln\frac{\eta+\zeta}{\eta-1}\big)-\frac{1}{1-\zeta^2}+\frac{\mu}{p^z(1-\eta)^{2}} & \eta<-\zeta \\
-f(-\zeta,\eta)\ln\frac{p_z^2}{\mu^2}-f(\zeta,\eta)\ln\frac{1-\eta}{\zeta-\eta}-f(-\zeta, \eta)\ln[4(\zeta+\eta)(1-\eta)]\non\\
+4f(-\zeta, \eta)-\frac{1}{1-\zeta^2}-\frac{1}{1+\zeta}+\frac{3}{1-\eta}+\frac{\mu}{p^z(1-\eta)^{2}} &-\zeta<\eta<\zeta \\
-(f(\zeta, \eta)+f(-\zeta, \eta))\ln\frac{p_z^2}{\mu^2}-f(\zeta, \eta)\ln[4(\eta-\zeta)(1-\eta)]\non\\
-f(-\zeta, \eta)\ln[4(\zeta+\eta)(1-\eta)]+4(f(\zeta, \eta)+f(-\zeta, \eta))+\frac{6}{1-\eta}\non\\
-\frac{3}{1-\zeta^2}+\frac{\mu}{p^z(1-\eta)^{2}} & \zeta<\eta<1 \\
-f(\zeta, \eta)\ln\frac{\eta-\zeta}{\eta-1}-f(-\zeta, \eta)\ln\frac{\eta+\zeta}{\eta-1}+\frac{1}{1-\zeta^2}+\frac{\mu}{p^z(1-\eta)^{2}} & \eta>1.
\end{cases}\\
\end{align}
One can check to see that $\delta Z_H^{(1)}$ provides a plus distribution for the singularity at $x=y$ in the matching factor $Z_H^{(1)}(x/y,\xi/y,\mu/p^z)$. The above matching factor also transforms the logarithmic dependence on $p^z$ in ${\cal H}(x,\xi, t, \mu, p^z)$  into the renormalization scale dependence in $H(x, \xi, t, \mu)$. Note that we have the same linear divergence as in the PDF case. Moreover, when $\xi\to 0$, the region $[-\xi,\xi]$ disappears and the matching factors in the remaining regions reduce to those for the PDF.

In the above result, we take into account the quark contribution only. The antiquark contribution is given by making the following replacement in Eq.~(\ref{1loopmatchingfull})
\beq\label{antiquarkrep}
x\to-x,\hspace{5em} y\to-y.
\eeq
Summing over both the quark and the antiquark contribution, one obtains the complete matching factor, which contains in the $\ln\frac{p_z^2}{\mu^2}$ term the complete evolution kernel of the light cone GPD.

The factorization of $E$ can be constructed analogously. However, from the results of $\cal E$ and $E$, the matching factor for $E$ is simply given by
\beq
Z_E(x/y)=\delta(x/y-1)
\eeq
up to one-loop order and leading $p^z$ accuracy. Since $E$ does not show up at tree level, it is UV convergent and thus does not have a cutoff dependence. Accordingly, $\cal E$ does not have a logarithmic dependence on $p^z$. Therefore, the light cone GPD $E$ can be smoothly approached by the large momentum limit of $\tilde E$. This is also true, in principle, for other light cone quantities that do not exhibit a UV divergence. The simulation of such quantities on the lattice are, therefore, relatively simple.

Since $\tilde {\cal H}^{(1)}={\cal H}^{(1)}$, $\tilde H^{(1)}=H^{(1)}$, and $\tilde E^{(1)}=\tilde {\cal E}^{(1)}$, the above factorization also applies to the polarized GPDs with the same matching factors.


In the following, we consider the matching for the distribution amplitude, whose evolution kernel can be obtained from that of the GPD as a limiting case. Here we focus on the simplest type, the distribution amplitude of the pion. The light cone pion distribution amplitude $\phi(x)$ is given by
\beq\label{LCDA}
\phi(x)=\int\frac{dz^-}{2\pi}e^{i(2x-1)p^+ z^-/2}\langle \pi(p)|\bar\psi(-\frac{z}{2})\gamma^+\gamma_5 L(-\frac{z}{2},\frac{z}{2})\psi(\frac{z}{2})|0\rangle,
\eeq
where the two quark fields are separated along the light cone, and $x\ (1-x)$ denotes the momentum fraction of the quark (antiquark). As in the case of GPDs, it can be studied from the large momentum limit of the following quasi correlation
\beq\label{quasiDA}
{\tilde \phi}(x, p^z)=\int\frac{dz}{2\pi}e^{-i(2x-1)p^z z/2}\langle \pi(p)|\bar\psi(-\frac{z}{2})\gamma^z\gamma_5 L(-\frac{z}{2},\frac{z}{2})\psi(\frac{z}{2})|0\rangle
\eeq
with the two quark fields separated along the spatial direction. The one-loop factorization for the pion distribution amplitude can be written down analogously as
\beq
{\tilde \phi}(x, \mu, p^z)=\int_0^1 dy\, Z_\phi(x, y, \mu, p^z)\phi(y, \mu).
\eeq
The matching factor $Z_\phi(x, y, \mu, p^z)$ can be obtained by starting with the light cone and quasi distribution amplitudes, Eqs.~(\ref{LCDA}) and (\ref{quasiDA}), and computing the one-loop corrections, respectively. It can also be obtained from the above matching factor for GPDs by crossing the initial quark to the final state, which leads to the following replacement for the momentum fractions
\beq
\zeta\to \frac{1}{2y-1}, \hspace{1em} \eta/\zeta\to 2x-1.
\eeq
This corresponds to setting
\beq\label{DArep}
x\to 2x-1, \hspace{5em} y\to 2y-1, \hspace{5em} \xi\to 1
\eeq
in Eq.~(\ref{1loopmatchingfull}). Note that one also needs to replace the averaged longitudinal momentum of external hadrons $P^z$ below Eq.~(\ref{1loopmatchingfull}) by $p^z/2$ with $p^z$ being the longitudinal momentum of the pion.

Expanding the matching factor $Z_\phi(x, y, \mu, p^z)$ as
\beq
Z_\phi(x, y, \mu, p^z)=\delta(x-y)+\frac{\alpha_S}{2\pi}Z_\phi^{(1)}(x, y, \mu, p^z)+\dots,
\eeq
e then have
\begin{align}\label{DAmatfac}
Z_\phi^{(1)}(x, y,\mu, p^z)/C_F&=G_1(x, y, \mu, p^z)\theta(x<0)+G_2(x, y, \mu, p^z)\theta(0<x<y)\non\\
&+G_3(x, y, \mu, p^z)\theta(y<x<1)+G_4(x, y, \mu, p^z)\theta(x>1)
\end{align}
with
\begin{align}\label{DAmatfac1}
G_1(x, y, \mu, p^z)&=\big(\frac{x}{2(1-y)}+\frac{1-x}{2y}\big)\ln\frac{x}{x-1}+\big(\frac{x}{2(1-y)}-\frac{1-x}{2y}+\frac{1}{y-x}\big)\ln\frac{(x-y)^2}{x(x-1)}\non\\
&+\frac{\mu}{p^z (x-y)^2}, \non\\
G_2(x, y, \mu, p^z)&=\big(\frac{x-1}{y}+\frac{1}{y-x}\big)\ln\frac{p_z^2}{\mu^2}+\big(\frac{x}{2(1-y)}-\frac{1-x}{2y}+\frac{1}{y-x}\big)\ln[4(x-y)^2]\non\\
&+\big(\frac{x}{2(y-1)}+\frac{x-1}{2y}\big)\ln[4x(1-x)]
+\big(\frac{x}{2(y-1)}+\frac{1-x}{2y}-\frac{1}{y-x}\big)\ln\frac{1-x}{x}\non\\
&+\frac{1}{y}-\frac{1}{y-x}+\frac{\mu}{p^z (x-y)^2}, \non\\
G_3(x, y, \mu, p^z)&=G_2(1-x, 1-y, \mu, p^z),\non\\
G_4(x, y, \mu, p^z)&=\big(\frac{x}{2(1-y)}+\frac{1-x}{2y}\big)\ln\frac{x-1}{x}+\big(\frac{x}{2(1-y)}-\frac{1-x}{2y}+\frac{1}{y-x}\big)\ln\frac{x(x-1)}{(x-y)^2}\non\\
&+\frac{\mu}{p^z (x-y)^2},
\end{align}
Here we have taken into account charge conjugation invariance and the fact that one-loop diagram for the pion simultaneously involves the quarks and the antiquarks state. We have checked to see the above matching factor agrees with the result of direct computation.

Near $x=y$, one has an extra contribution from the wave function renormalization
\beq
Z_\phi^{(1)}(x, y,\mu, p^z)/C_F=\delta Z_\phi^{(1)}(2\pi/\alpha_S)\delta(x-y),
\eeq
where $\delta Z_\phi^{(1)}$ is given by
\begin{align}\label{DAmatfacdeltaZ}
\delta Z_\phi^{(1)}&=-\frac{\alpha_S C_F}{2\pi}\int dx\begin{cases}
(\frac{x-1}{2(1-y)^2}+\frac{1}{2(y-1)}-\frac{1}{y-x})\ln\frac{1-x}{y-x}+(\frac{x}{2y^2}+\frac{1}{2y}-\frac{1}{y-x})\ln\frac{x}{x-y}\non\\
-\frac{1}{2(y-1)}-\frac{1}{2y}+\frac{\mu}{p^z(y-x)^2} & x<0\\
(-\frac{x}{2y^2}-\frac{1}{2y}+\frac{1}{y-x})(\ln\frac{p_z^2}{\mu^2}+\ln[4x(y-x)])+(\frac{1-x}{2(1-y)^2}-\frac{1}{2(y-1)}\non\\
+\frac{1}{y-x})\ln\frac{x-y}{x-1}-\frac{1}{2(y-1)}+\frac{2x}{y^2}+\frac{1}{2y}-\frac{1}{y-x}+\frac{\mu}{p^z(y-x)^2} & 0<x<y \\
(-\frac{1-x}{2(1-y)^2}-\frac{1}{2(1-y)}+\frac{1}{x-y})(\ln\frac{p_z^2}{\mu^2}+\ln[4(1-x)(x-y)])+(\frac{x}{2y^2}\non\\
+\frac{1}{2y}-\frac{1}{y-x})\ln\frac{x-y}{x}+\frac{1}{2y}+\frac{2(1-x)}{(1-y)^2}+\frac{1}{2(1-y)}+\frac{1}{y-x}+\frac{\mu}{p^z(y-x)^2} &y<x<1\\
-(\frac{x-1}{2(1-y)^2}+\frac{1}{2(y-1)}-\frac{1}{y-x})\ln\frac{1-x}{y-x}-(\frac{x}{2y^2}+\frac{1}{2y}-\frac{1}{y-x})\ln\frac{x}{x-y}\non\\
+\frac{1}{2(y-1)}+\frac{1}{2y}+\frac{\mu}{p^z(y-x)^2} & x>1.
\end{cases}\\
\end{align}
One can explicitly check to see that the wave function renormalization factor provides a plus prescription for the factor in Eq.~(\ref{DAmatfac}).

\section{Conclusion}
We have presented the one-loop matching conditions for the unpolarized and polarized generalized quark distribution in the nonsinglet case. The matching conditions relate the quasi GPDs defined in terms of spacelike correlations and the light cone GPDs. For the GPD $H$ ($\tilde H$), the matching is constructed in analogy with the PDF matching, and the matching factor reduces to that of the PDF in the limit $\xi\to 0$. For $E$ ($\tilde E$), as it is UV convergent, the matching factor is trivially given by a $\delta$ function, implying that the light cone GPD $E$ ($\tilde E$) can be smoothly approached by its quasi counterpart $\cal E$ ($\tilde {\cal E}$) in the large momentum limit. This facilitates its extraction from lattice simulations. We have also presented the matching condition for the pion distribution amplitude.
\section{Acknowledgments}
We thank V. Braun for the useful discussions on the pion distribution amplitude, and M. Diehl for the helpful communications on GPDs and distribution amplitudes. This work was partially supported by the U. S. Department of Energy via grants DE-FG02-93ER-40762, by grant No. 11DZ2260700) from the Office of Science and Technology in Shanghai Municipal Government, and by the National Science Foundation of China (Grants No. 11175114 and No. 11405104), and a DFG grant Grant No. SCHA 458/20-1.

\end{document}